# Joint Milli-Arcsecond Pathfinder Survey (JMAPS): Overview and Application to NWO Mission

B. Dorland and R. Dudik

US Naval Observatory

11 March 2009

## 1 Mission Overview

The Joint Milli-Arcsecond Pathfinder Survey (JMAPS) mission is a Department of Navy (DoN) space-based, all-sky astrometric bright star survey. JMAPS is currently funded for flight, with at 2012 launch date. JMAPS will produce an all-sky astrometric, photometric and spectroscopic catalog covering the magnitude range of 1—12, with extended results through 15$^{th}$ magnitude at an accuracy of 1 milliarcsecond (mas) positional accuracy at a mean observing epoch of approximately 2013. Using Hipparcos and Tycho positional data from 1991, proper motions with accuracies of 100 microarcseconds (umas) per year should be achievable for all of the brightest stars, with the result that the catalog will degrade at a much reduced rate over time when compared with the Hipparcos catalog.

JMAPS will accomplish this with a relatively modest aperture, very high accuracy astrometric telescope flown in low earth orbit (LEO) aboard a microsat. Mission baseline is for a three-year mission life (2012—2015) in a 900 km sun synchronous terminator orbit.

# 2 Mission Status

## 2.1 Programmatic Overview

With the release of Program Decision Memorandum-III (PDM-III) in November 2008, JMAPS is fully funded for flight through 2015. The resource sponsor is the Oceanographer of the Navy, with primary execution responsibilities assigned to the Office of Naval Research (ONR). The US Naval Observatory (USNO) has Principle Investigator (PI) and data processing responsibilities. The Naval Research Laboratory (NRL) has responsibility for the space, launch, ground station and mission operations center segments.

## 2.2 Schedule

Based on the current funding profile, JMAPS is scheduled for System Requirements/System Design review in calendar year 2009, Preliminary Design Review in 2010, Critical Design and Technology Readiness Review in 2011, and launch in 2012. All technologies will be at TRL 6 by the end of calendar year 2009. The mission will operate from 2012—2015, with a final astrometric and photometric catalog release in 2016.

# 3 Output products

There are two relevant output products, or deliverables, from JMAPS: first, JMAPS will generate an astrometric and photometric star catalog, accurate to 1 mas or better through $12^{th}$ magnitude. Second, JMAPS will mature, demonstrate and fly attitude sensing instrumentation that is accurate to 5 mas. In addition to maturing large format CMOS-Hybrid FPA, high-stability SiC optics, and advanced electronics technology, JMAPS will demonstrate the ability to build and operate 10-mas class attitude sensors, all by 2012. The catalog, technology and instrumentation demonstration are all highly relevant to future NASA missions such as NWO. In the following sections, we describe instrument design in more detail, and discuss how JMAPS might have direct application to NWO.

# 4 Spacecraft overview

As show in fig. 1, JMAPS is a single-instrument spacecraft consisting of the instrument deployed on a payload deck that is affixed to the top of the bus. The instrument/payload deck consists of the optical telescope assembly (OTA), the focal plane assembly (FPA), the instrument electronics (i.e., processor and memory), thermal control for the FPA, a sun shield, and the coarse star tracker. The instrument component occupies a volume 16.2" (41.1 cm) (h) x 25.8" (65.5 cm) x 24" (61.0 cm) with a mass of 48 kg, including 25% margin.

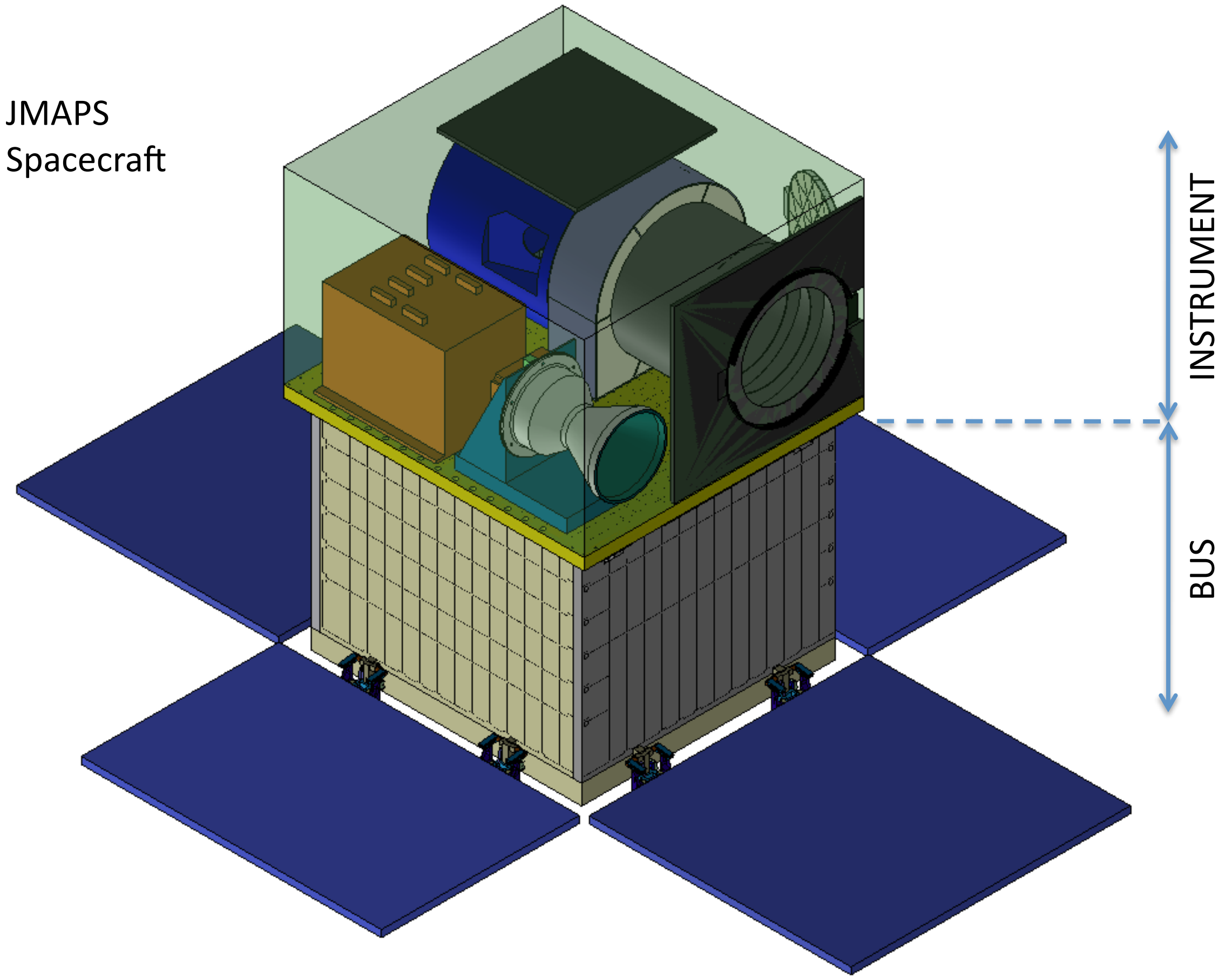

**Figure 1. JMAPS Spacecraft**

The lower portion of the spacecraft consists of the bus, which supports the solar panels and houses the power, avionics, communications, primary thermal control, reaction wheels, and inertial measurement units (IMUs). The total mass of the spacecraft, including contingency, is approximately 180 kg. The spacecraft occupies a volume approximately 38” (96.5 cm) (h) x 28” (71 cm) x 24” (61 cm).

In the JMAPS mission design, the Attitude Determination and Control System (ADCS) is integrated between the instrument and the bus. While slewing or in acquisition mode, coarse attitude information at the 1 arcsecond level is generated by the star tracker. This information is combined with data from the inertial reference unit (IRU) in the bus and processed with to generate pointing knowledge at or below the 1 mas accuracy level. During standard observations, the primary instrument is used to observe reference stars and generate boresight pointing quaternions at or below 10 mas in accuracy at 5 Hz. The 10 mas attitude signal is used by the ADCS system to drive the spacecraft pointing. Using a system of reaction wheels the spacecraft achieves pointing control and stability of 50 mas.

The primary relevance of this to the NWO program is that the JMAPS attitude determination subsystem—that is, the components of JMAPS that generate

quaternions—are contained in the instrument. The JMAPS instrument can be separated from the bus and, in principle, used on other platforms for very high accuracy attitude determination. This assumes that the requisite thermal and power connections are available. The accuracy of resultant attitude determination will be a function of platform considerations such as orbit, thermal stability, vibrational stability, etc.

## 5 Instrument Description

Figure 2 shows the layout of the instrument on the payload. The major components are all shown, i.e.: the OTA, the IEB, the star tracker and the FPA radiators. Not shown are the sun shield (deployed on the rear of the payload deck to protect the OTA from direct exposure to the sun), thermal insulation around the OTA, the loop heat pipe structure to carry heat to the radiators, or the structure associated with supporting the radiators. Also not visible is the FPA . The FPA is contained inside the OTA, located on the payload deck in relatively close proximity to the IEB. In the follow subsections, additional details are provided about the major instrument subsections.

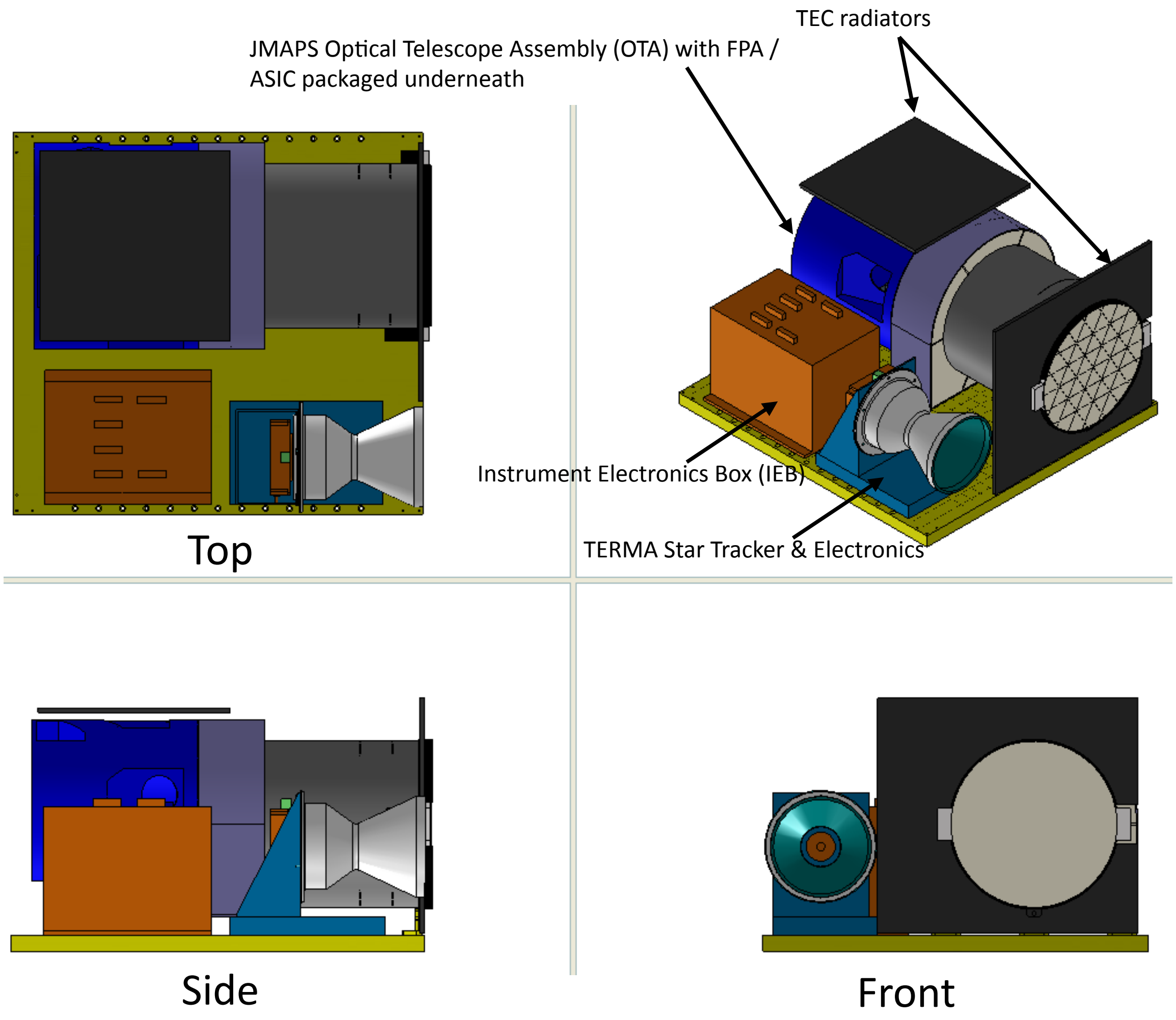

Figure 2 JMAPS Instrument—conceptual design of payload deck.

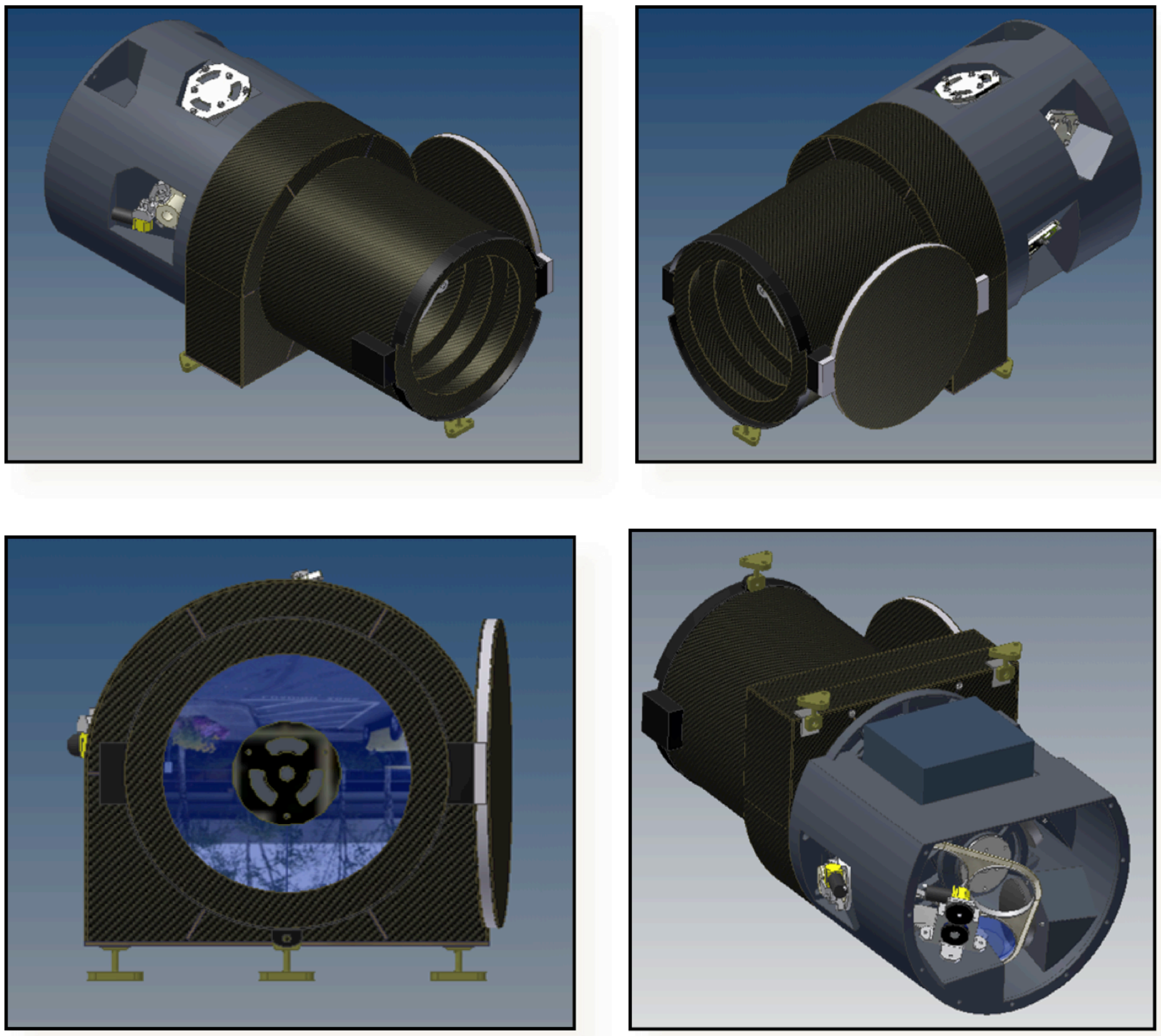

Figure 3 Various views of the JMAPS Optical Telescope Assembly showing housing, metering structure, optics, baffling and cover. (Bottom right) cutaway view shows the filter wheel (rear) and FPA housing (top) of OTA as viewed from below.

The JMAPS instrument has been designed to meet a 5 mas single measurement systematic error floor, including both local (i.e., centroiding), differential (e.g., residual distortion), and detector error sources. The design and system engineering processes required to meet this specification are currently underway and include high-fidelity thermal, vibrational, optical and focal plane modeling. Results of these design and analyses activities will be reported out at appropriate venues and publications.

## 5.1 Optical Telescope Assembly

The high precision and repeatability of the required measurements necessitates an optical system with very low aberrations and very high stability. The JMAPS optical telescope is a 19 cm, f/20, on-axis space astrograph designed for a residual aberration and distortion floor of below 5 mas. This is achieved through the use of Silicon Carbide (SiC) optics and metering structure for the powered optics, and through a thermal design that maximizes the thermal stability of the optical elements and structure.

The OTA consists of a plane-parallel entrance aperture window, made of fused silica. This window provides both a red band cutoff at about 920 nm as well as mechanical support for M2, with minimal stray light effects. There are three powered elements and six fold flats that are used to contain the 3.8 m effective focal length in a volume that is approximately 25" along the long axis. The system includes a rotating filter wheel with four positions. The first position is a wide-open visible band (~400-900 nm), the second position is the primary astrometric bandpass blue cutoff (it passes light > 700 nm), the third is a low resolution spectral grating (R~15) used for basic color sensing, and the fourth is TBD.

The design residual wave front error (WFE) is well below 0.05 waves at 633 nm, while the as-built WFE is between 0.05 and 0.067 waves at 633 nm. Residual centroiding error due to static (e.g., PSF undersampling) and varying (e.g., thermally-induced defocus) sources of error will not exceed 3 mas RMS. Residual distortion and other sources of differential astrometry error due to both static and varying effects will likewise be kept below 3 mas. These sources of error roll up to 4 mas for the optics-induced single-measurement JMAPS systematic floor contribution.

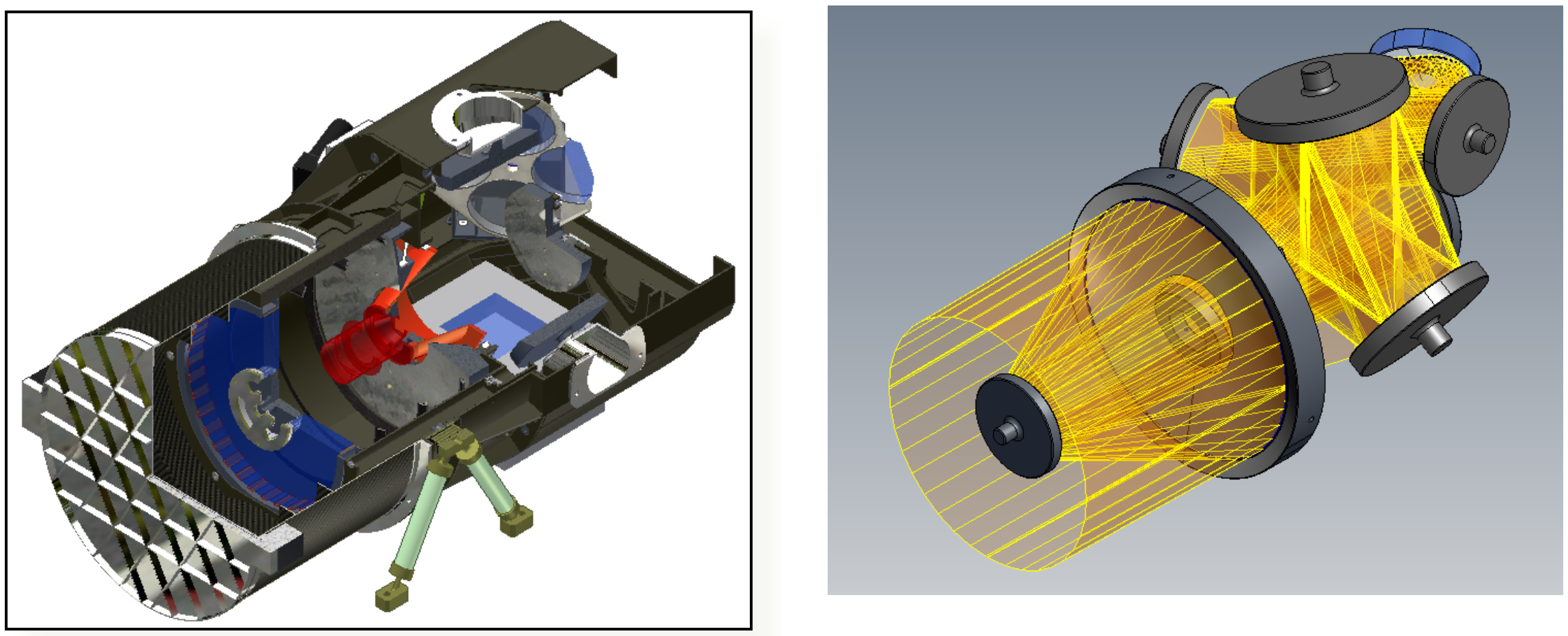

**Figure 4: (left) JMAPS Internal Optical Telescope Assembly (OTA) and (right) optical raytrace**

## 5.2 Focal Plane Assembly

The baseline JMAPS focal plane assembly (FPA) consists of a 2 x 2 mosaic of Teledyne Imaging Sensor (TIS) H4RG-10 sensor chip assemblies (SCAs), as shown in Figure 5. The H4RG-10 is a 4192 x 4192, 10 micron pixel CMOS-Hybrid SCA. JMAPS will use the B0, or 3$^{rd}$ generation design of the SCA. The first generation SCA (A1) has been fabricated and tested at both Goddard Space Flight Center (GSFC) and USNO and reported on in the literature. USNO will take delivery of the second generation SCA (A2) in 2009 and immediately begin laboratory, sky and radiation testing working with both GSFC and the Air Force Research Laboratory (AFRL). By early summer, initial results will be available for the A2 design. At this point, the A2 design will have achieved a TRL of 6.

The flight SCAs will be mounted on an integrated FPA that will provide electric, thermal and mechanical interfaces to the packaged SCAs. The FPA will be cooled to a temperature of approximately 193 K, with a stability specification of 10 mK. The FPA heat will be dumped to space using dedicated FPA radiators on the front aperture and above the OTA (see Figure 2).

The FPA is allocated 3 mas of systematic error for centroiding measurements. This includes the effects of charge diffusion, intra- and inter-pixel quantum efficiency variations, hot pixels, inter-pixel electronic cross-talk, and SCA motion. These requirements drive the performance specifications for the flight units as well as the thermal and mechanical stability requirements on the FPA. FPA requirements and performance are currently being validated using a number of tools, including a high-fidelity focal plane simulator.

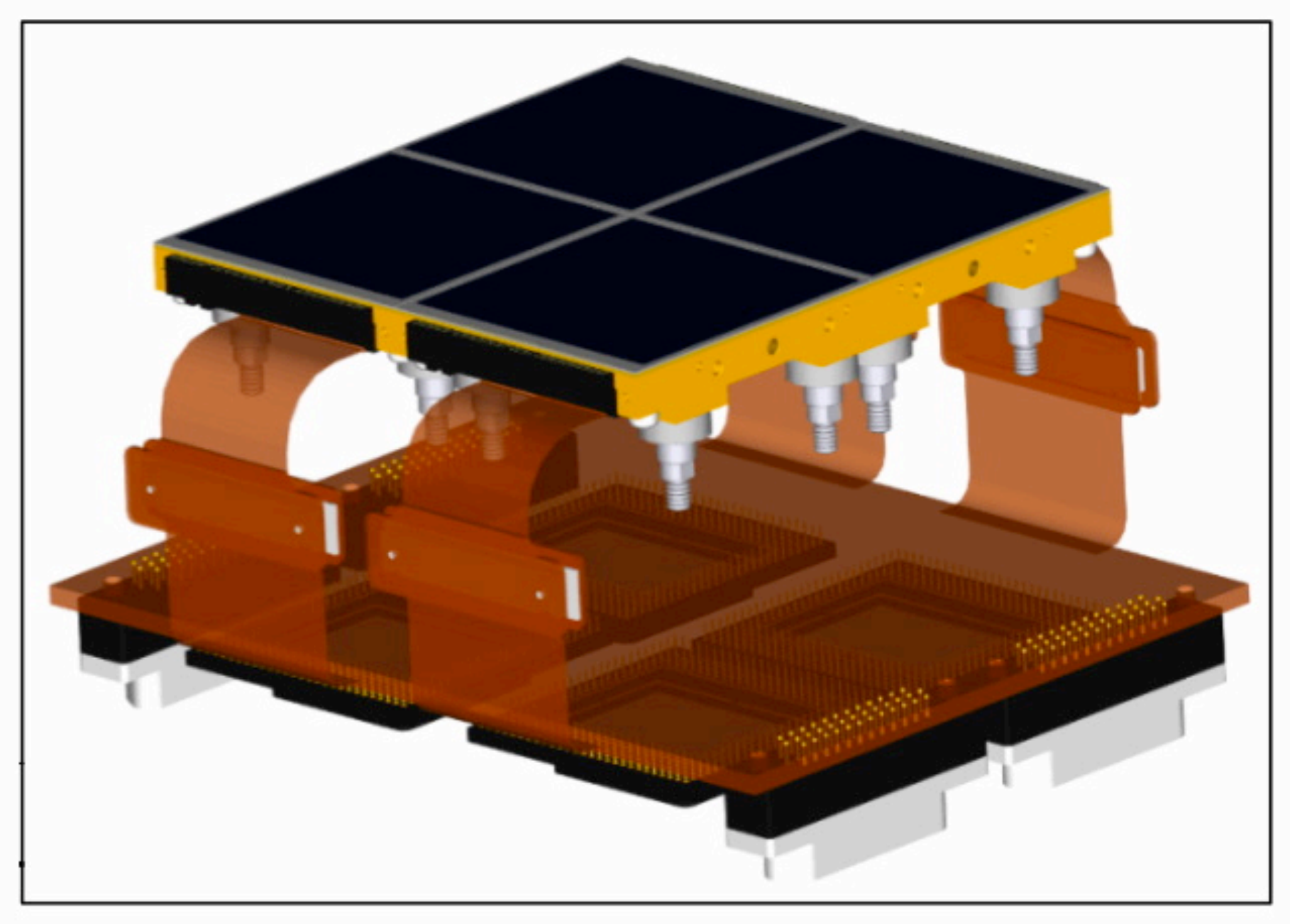

**Figure 5: JMAPS FPA 4K 2X2 Mosaic Concept**

## 5.3 Readout ASICS

The baseline JMAPS readout and FPA control electronics are the TIS SIDECAR ASICS (see Figure 6). JMAPS will utilize a total of four to operate the FPA. The SIDECARS digitize the data coming off the FPA, with a 16-bit depth. Multiple output connections will be used to readout the FPA in approximately 1 second. In addition, for both guide star and bright star applications, the SIDECARs will readout smaller windows on the FPA at higher frame rates. Guide star (approximately 10 stars per field of view) windows will be read out at 5 frames per second (fps), while bright star (< 5th magnitude) windows will be read out at frame rates of up to 200 fps.

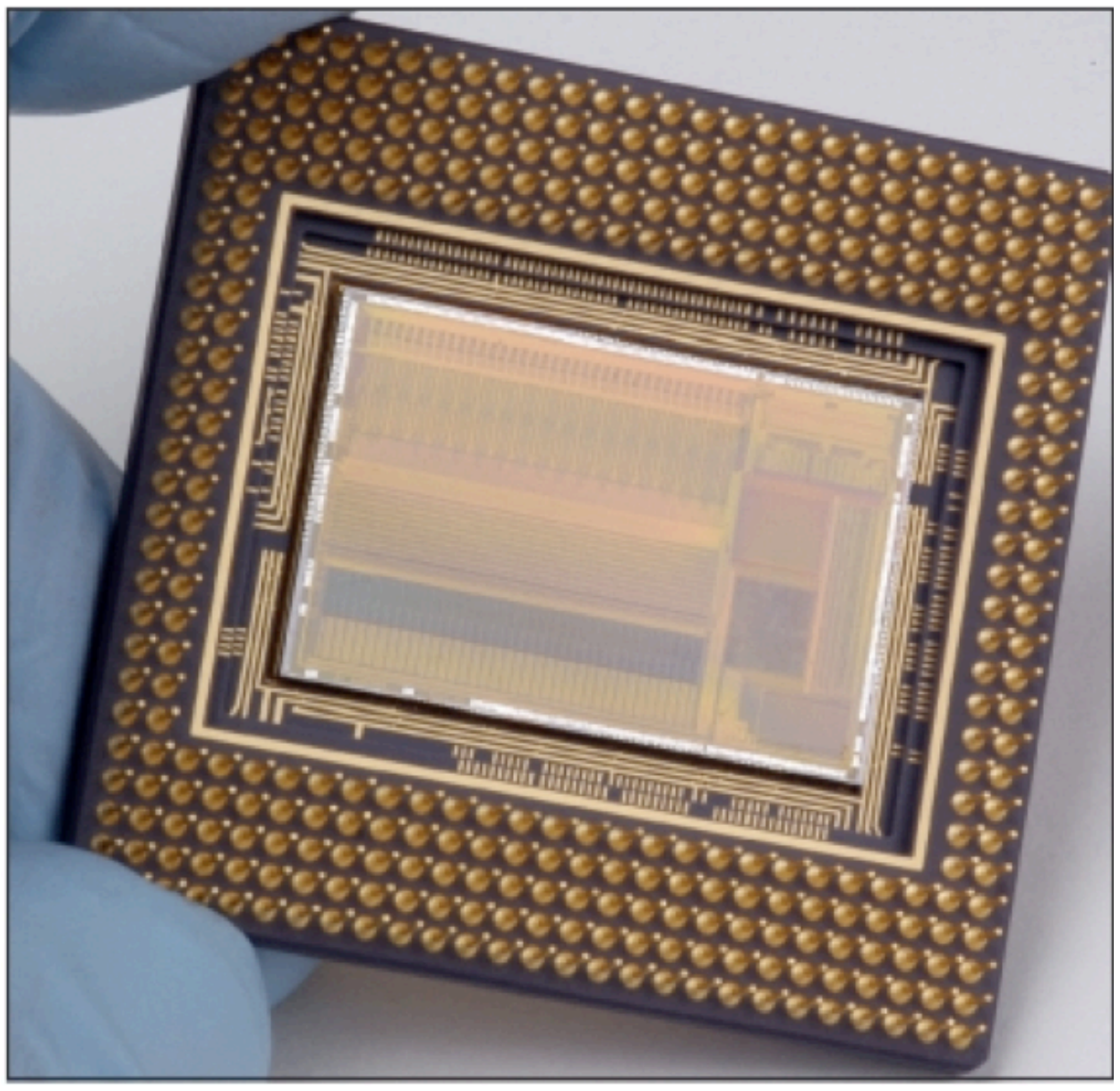

Figure 6: Sidecar ASIC

## 5.4 Instrument Electronics Box

The Instrument Electronics Box (IEB) houses the primary on-board processing system for the instrument and spacecraft. The IEB combines the power conversion, mechanism and thermal control, data processing and data storage functions. The IEB receives full-frame data from the ASICS. CDS processing is performed by differencing science observations from the reset readout frame. The processor then steps through the image and detects stars. For each detected star, a 10 x 10 pixel windows is retained and stored in the Solid State Data Recorder (SSDR) for later downlink to the ground. Attitude determination calculations are performed in the IEB, and quaternions are generated to drive the ADCS. Data rates for the windows are typically about 5.6 Mbits per minute. Data are downlinked to the ground station via the on-baord transmitter during regular passes. One pass per day over the ground station is sufficient to download one day's worth of astrometric data.

## 5.5 Integrated Instrument Concept of Operations and Predicted Performance

The JMAPS instrument is operated in a fashion similar to standard star trackers. A star field is imaged—in the case of JMAPS, integration times of 1, 4.5 and 20 seconds are used—individual star positions are calculated, and the boresight pointing calculated every 200 msec. As described in Table 1, the integrated instrument has a field of view of 1.24° x 1.24°. Each pixel subtends approximately 0.5 seconds of arc. The primary astrometric bandpass is 700—900 nm, approximately equivalent to astrometric I band. The typical point spread function (PSF) full width at half maximum (FWHM) is 0.87 arcseconds, for a sampling of 1.6 pixels per FWHM.

Using a combination of focal plane and centroiding models combined with analyses of the PSFs, instrument throughput, etc., we can predict centroiding accuracy for single measurements as a function of I band magnitude. These results are shown in Figure 7. The results consider only the "best case" for a discrete range of exposure times (0.01, 0.2, 1.0, 4.5 and 20 seconds). Here, best case is defined as the longest, non-saturated exposure time.

The ADCS system uses the ten brightest stars on the focal plane to calculate a boresight attitude at a 5 Hz rate. Using typical numbers and I–band brightnesses, boresight accuracies of better than 10 mas at a 5 Hz rate can be achieved according to JMAPS attitude determination simulation results.

The instrument performance estimates are based on detailed simulations of the focal plane, optics, centroiding process and a realistic attitude determination system simulator. The results include both thermal and vibrations effects.

| Parameter | Spec | Units |
|---|---|---|
| **Optics** | | |
| Optical Aperture | 19 | cm |
| Collecting Area | 223.00 | cm^2 |
| central obscuration | 9 | cm |
| Focal length. | 3.8 | m |
| Focal Ratio | 19.900 | |
| Optical Field of View (FOV) | 1.24 | square degrees |
| Survival Temperature | 180-320 | K |
| Operating Temperature | 200-240 | K |
| Single Measure Instrument Error Floor | 5 | mas |
| Mission Systematic Error Floor | 0.5 | mas |
| **Focal Plane** | | |
| Size of FPA (pixels) | 8192 | square pixels |
| Size of FPA (mm) | 81.92 | square mm |
| Pixel size | 10 | um |
| Pixel subtense | 0.545 | arcseconds |
| FPA Read noise | 10 | e- |
| Mode of dark current distribution | <1 | e/s/pix |
| Operating Temperature. | 193 | K |
| Survival Temperature | 180 | K |
| Temperature stability | < 0.01 | K |
| Spectral Range | 450 - 900 | nm |
| Full Well | 100000 | e- |
| Residual uncorrected detector effects | < 3.5 | mas |
| **Instrument** | | |
| Data Rate | | |
| Per Day | 93 | kbits/sec |
| Per Minute | 5.6 | Mbits/min |
| Per Day | 7.5 | Gbits/day |
| FPA Readout time | ≤ 1 | seconds |
| Digitization noise | ≤10 | e- |
| Overall instrument read noise | ≤14 | e- |

**Table 1. Instrument performance parameters (notional).**

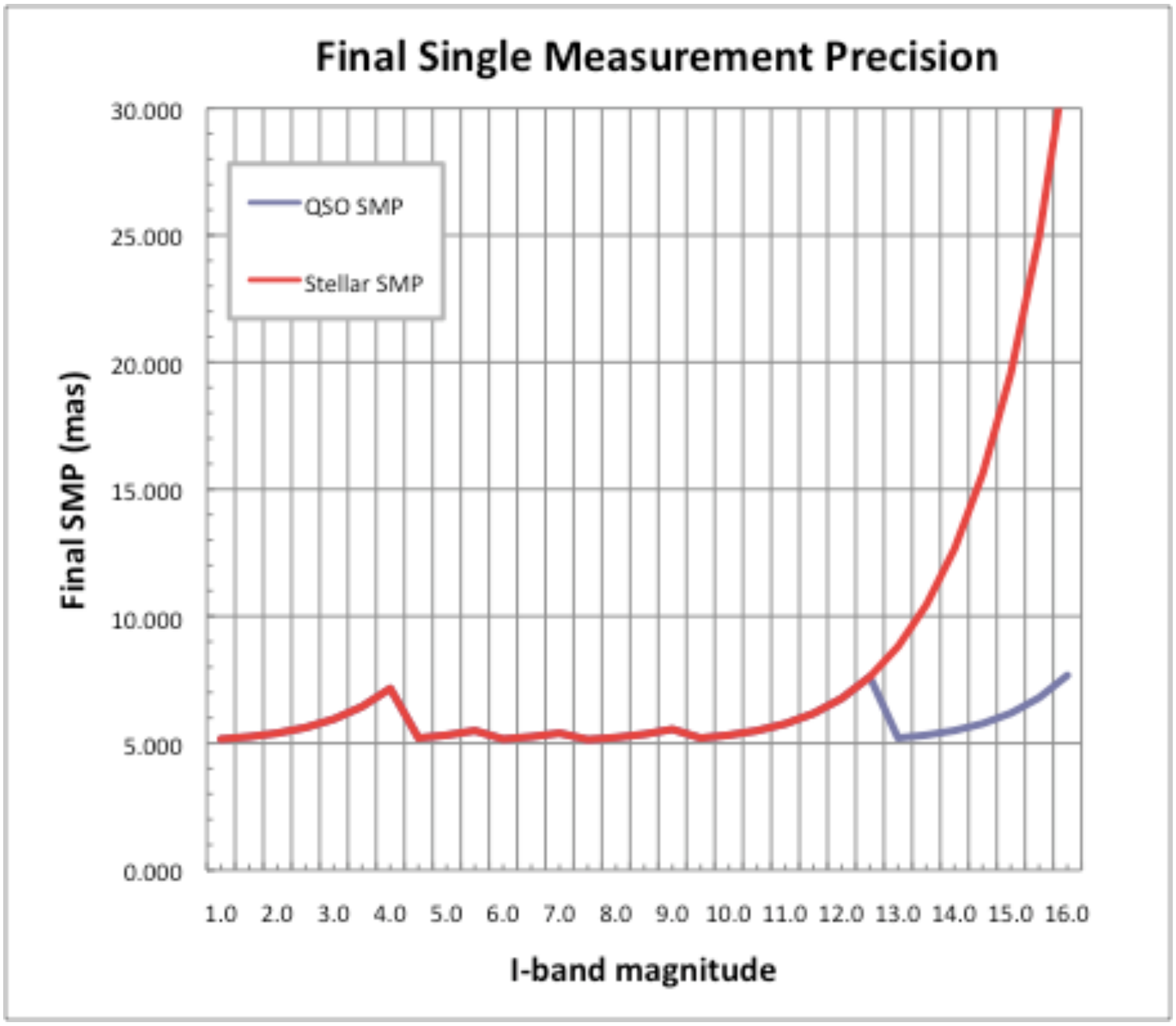

Figure 7. Single measurement precision predictions for JMAPS as a function of souce I-band magnitude.

# 6 NWO-Specific Applications

## 6.1 Similarities and Differences between baseline JMAPS and NWO implementation

The description provided in §4 and 5 refer to the baseline JMAPS mission. This mission will fly in the 2012—2015 timeframe. By 2013 all technology and concepts will be demonstrated at TRL 9 and by 2016 a final catalog will be released with stellar positions accurate at the 1 mas level.

As noted in §5, the instrument is essentially a discrete component and can be thought of as an extremely accurate attitude sensor/star tracker. A second instance of the instrument can be built and deployed, in principle, on other platforms. Assuming the proper mechanical, thermal and electronic interface, it can then serve as an extremely accurate star sensor. If flown after 2016, it will also be able to take advantage of the JMAPS star catalog. Based on extensive cost modeling using both the NASA Instrument Model (developed by JPL) and Aerospace Corporation's Concept Design Center, the instrument development cost is estimated at under $40M ($FY08). A significant portion of this is non-recurring expenses, so a second unit would likely cost $30M or less. Adopting a $40M not-to-exceed limit would be a very conservative estimate, and would allow for some modifications to be made based on lessons learned with the first unit.

For purposes of the NWO analysis, we assume that a second instrument is built and deployed as star sensor onboard the NWO starshade. The instrument would be used to navigate the starshade during the alignment phase of the occultation. The original instrument is designed to support 5 mas single measurement accuracy for a LEO-class mission. Most of the sources of systematic error are related to varying thermal conditions over the orbit. A similar unit deployed at L2 would be in a significantly more benign thermal environment. As such, we would expect the systematic error floor to be well below 5 mas. For purposes of this analysis, however, we will continue to use the 5 mas value from the baseline mission. We will continue the analysis for the L2 case and will publish these results in the appropriate forum once the analysis is completed.

## 6.2 Starshade alignment

The primary application of the JMAPS instrument to NWO is to support the alignment of the starshade (see Figure 8). The overall process is discussed in detail elsewhere in the main body of the NWO report; we restrict our description here to the JMAPS component. In the case of disagreement between the description presented here and that presented elsewhere in the NWO report, the latter takes precedence over this one.

In order to suppress the starlight and reveal the reflected light from the exoplanets, the NWO starshade has to be accurately maneuvered in front of the telescope at distances of 50—100 Megameters (50,000—100,000 km). The JMAPS instrument will be deployed on the starshade and oriented towards the telescope. In order to ensure the required signal, an I-band beacon with a maximum apparent magnitude of 12 at the maximum alignment distance will be affixed to the telescope and directed at the starshade.

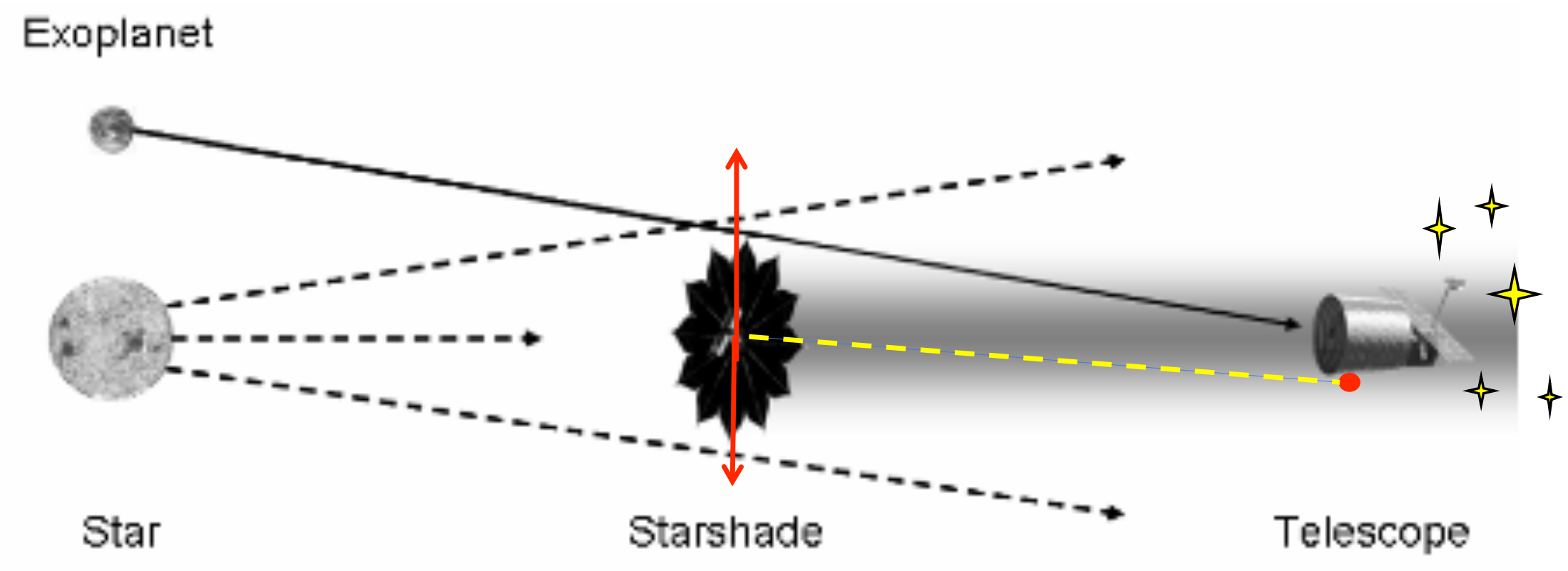

Figure 8: Illustration of NWO alignment problem

The JMAPS instrument will observe the telescope beacon against the background field of reference stars. Assuming a typical background guiding signal update rate of 0.1 Hz, the JMAPS instrument will average the boresight pointing results over 50 observations, and generate a mean boresight measurement of accuracy 1 mas.

If we adopt a 10 second integration time, we can calculate the single measurement accuracy for individual, stellar-like sources. Results are presented for three typical NWO ranges in Table 2. The nominal NWO 12$^{th}$ magnitude beacon is highlighted in the table. The conclusion is that the instrument is capable of 8 mas position determination given a ten second exposure. When combined with the boresight accuracy of 1 mas over the same period, alignment accuracies (1-σ) of between 2 and 4 meters are expected (see the table). The instrument will perform both the position and boresight calculations internally and generate boresight and telescope positions for the Starshade spacecraft once every ten seconds.

Increased exposure time, a brighter beacon and a reduced systematic error floor due to the more benign thermal environment at L2 would drive down the error. In principle, alignment errors of 1 meter may be possible, but additional analysis—including detailed thermal simulation—are required to support this result.

| Magnitude (I band) | SMP (mas) tint = 10.0 | Range (megameters) 50 | 80 | 100 |
|---|---|---|---|---|
| **0** | 5.00007 | 1.21 | 1.94 | 2.42 |
| **0.5** | 5.00010 | 1.21 | 1.94 | 2.42 |
| **1** | 5.00016 | 1.21 | 1.94 | 2.42 |
| **1.5** | 5.00026 | 1.21 | 1.94 | 2.42 |
| **2** | 5.00041 | 1.21 | 1.94 | 2.42 |
| **2.5** | 5.00065 | 1.21 | 1.94 | 2.42 |
| **3** | 5.00104 | 1.21 | 1.94 | 2.42 |
| **3.5** | 5.00164 | 1.21 | 1.94 | 2.42 |
| **4** | 5.00260 | 1.21 | 1.94 | 2.42 |
| **4.5** | 5.00412 | 1.21 | 1.94 | 2.42 |
| **5** | 5.00653 | 1.21 | 1.94 | 2.42 |
| **5.5** | 5.01035 | 1.21 | 1.94 | 2.43 |
| **6** | 5.01639 | 1.21 | 1.94 | 2.43 |
| **6.5** | 5.02595 | 1.22 | 1.95 | 2.43 |
| **7** | 5.04106 | 1.22 | 1.95 | 2.44 |
| **7.5** | 5.06492 | 1.23 | 1.96 | 2.45 |
| **8** | 5.10252 | 1.23 | 1.98 | 2.47 |
| **8.5** | 5.16154 | 1.25 | 2.00 | 2.50 |
| **9** | 5.25372 | 1.27 | 2.03 | 2.54 |
| **9.5** | 5.39662 | 1.31 | 2.09 | 2.61 |
| **10** | 5.61568 | 1.36 | 2.17 | 2.72 |
| **10.5** | 5.94641 | 1.44 | 2.30 | 2.88 |
| **11** | 6.43603 | 1.56 | 2.49 | 3.12 |
| **11.5** | 7.14394 | 1.73 | 2.77 | 3.46 |
| **12** | 8.14151 | 1.97 | 3.15 | 3.94 |
| **12.5** | 9.51219 | 2.30 | 3.68 | 4.60 |
| **13** | 11.35382 | 2.75 | 4.40 | 5.50 |
| **13.5** | 13.78423 | 3.34 | 5.34 | 6.67 |
| **14** | 16.95051 | 4.10 | 6.56 | 8.20 |
| **14.5** | 21.04207 | 5.09 | 8.15 | 10.18 |
| **15** | 26.30956 | 6.37 | 10.19 | 12.73 |
| **15.5** | 33.09343 | 8.01 | 12.81 | 16.02 |
| **16** | 41.87038 | 10.13 | 16.21 | 20.27 |

**Table 2. Predicted single measurement accuracies for stellar sources over typical NWO ranges. Results are presented in units of meters.**

## 6.3 Navigation

A secondary application of JMAPS to NWO is spacecraft navigation. As shown in Figure 9, the JMAPS instrument can be used to observe Solar system objects. By combining observations of multiple objects, the position of the instrument within the solar system can be determined. This is analogous to work currently be done in the Department of Defense which uses objects in orbit around the Earth to determine positions of sensors in Earth's vicinity. In this case, we are using much more distant Solar System objects to determine position within the Solar System.

An initial feasibility analysis has been performed. Possible targets include the outer planets, their satellites and minor planets. Key sources of error include: how to determine "position" for resolved targets such as the gas giants, the errors associated with the ephemerides of outer solar system objects, the signals for various candidate targets, and the effects of photocenter motion on sub-pixel centroiding. Initial analysis indicates that with sufficient signal, 1/100$^{th}$ pixel centroiding on quasi-point sources such as asteroids, and few AU baselines, positions with error ellipse axes of order 50 km or less are feasible. More detailed analysis is underway to refine a target list, develop an approach for out-of-plane position and assess accuracies with more fidelity. We will likely perform navigation experiments with the baseline JMAPS mission while on orbit in order to definitively determine the feasibility of this approach for Solar System navigation.

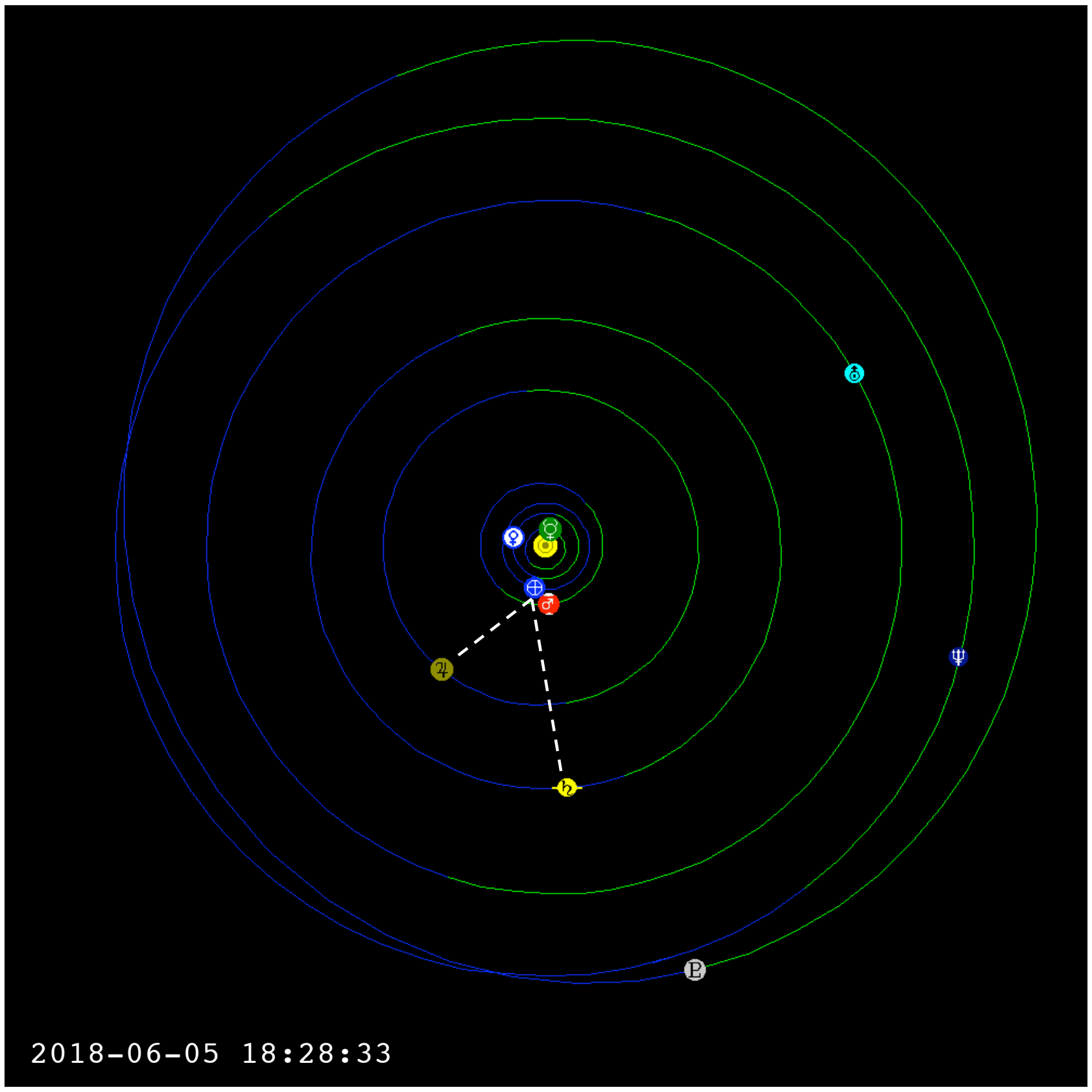

**Figure 9: Simulation of solar system orbits. JMAPS can be used to determine the position of the occultor to accuracies of 50km or less.**

# 7 Summary

The JMAPS mission is a funded, Department of Navy all-sky astrometric and photometric survey mission scheduled for launch in the 2012 timeframe. The mission will deliver an updated bright star catalog by 2016. The mission will develop and fly a new class of astrometric instrumentation that will support measurement of individual star positions at the 5 mas level and attitude determination at the 10 mas level at a 5 Hz rate. The instrument, described in detail in this paper, can be deployed on other platforms and used for both boresight attitude determination (i.e., attitude sensing) and individual star position measurement (i.e., star tracking). Initial analysis indicates that the catalog and technology can be applied to the NWO mission to support both alignment of the Starshade spacecraft at the 2—4 m accuracy level at distances of up to 100 Mm, and for Solar System navigation, with error ellipses of 50 km or less.